\documentstyle[fleqn,espcrc2_my,epsf]{article}

\def\lsim{\mathrel{\raise.3ex\hbox{$<$\kern-.75em\lower1ex\hbox{$\sim$}}}}
\def\gsim{\mathrel{\raise.3ex\hbox{$>$\kern-.75em\lower1ex\hbox{$\sim$}}}}
\def\Li2{{\rm Li}_2}

\newcommand{\bmE}{\mathbf E}

\newcommand{\bmx}{\mathbf x}
\newcommand{\bmy}{\mathbf y}

\newcommand{\AmS}{{\protect\the\textfont2
  A\kern-.1667em\lower.5ex\hbox{M}\kern-.125emS}}

\title{Delocalized Operator Expansion}

\author{A. H. Hoang\address{
   Max-Planck-Institut f\"ur Physik, 
   F\"ohringer Ring 6, 80805 M\"unchen, Germany
  }\thanks{Talk presented at QCD 02, Montpellier, France, July 2-9, 2002,
  and RadCor 2002, Kloster Banz, Germany, Sept. 8-13, 2002.}%
        }

\begin{document}




\begin{abstract}
A generalization of Wilson's local OPE for the short-distance expansion
of Euclidean current correlators, called delocalized operator
expansion (DOE), which has been proposed recently, is discussed. The
DOE has better convergence properties than the OPE
and can account for non-local non-perturbative QCD effects.
\end{abstract}

\maketitle

\section{INTRODUCTION - WILSON'S OPE} 

The Wilson OPE is one of the standard tools in modern hadronic
physics. The OPE provides the framework for systematically
separating short-distance contributions ($x\ll \Lambda^{-1}$)\footnote{  
I generically denote the low-energy hadronic scale by $\Lambda$.
}
from long-distance contributions ($x\sim\Lambda^{-1}$). 
For illustration consider the large-momentum expansion of the
correlator of a gauge-invariant QCD current $j(x)$~\cite{SVZ},
\begin{eqnarray}
\label{Wilson}
\lefteqn{
i\int\!d^4x e^{iqx}\langle T j(x)j^\dagger(0)\rangle
\stackrel{Q^2=-q^2\to\infty}{\longrightarrow}
} 
\nonumber\\[-2mm] &&
\mbox{\hspace{13mm}}
Q^2\sum_{N=0}\sum_{l=1}^{l_N}\,c_{Nl}(Q^2)\,\langle O_{Nl}(0)\rangle
\,
\end{eqnarray}
where $N$ runs over the dimension of the {\em local} and
gauge-invariant composite operators $O_{Nl}(0)$, and $l$ labels
operators of the same dimensions. The long-distance fluctuations are
encoded in the matrix elements $\langle O_{Nl}(0)\rangle$, also called
condensates, and the 
short-distance contributions are contained in the Wilson coefficients  
$c_{Nl}(Q^2)$. One has the scaling 
$\langle O_{Nl}(0)\rangle\sim\Lambda^N$ and $c_{Nl}(Q^2)\sim
Q^{-N}$. The $N=0$ term is entirely perturbative and higher order
terms in the series are of order $(\Lambda/Q)^N$. For the case
$Q\gg\Lambda$ the expansion should be reasonably
well-behaved, but it is known to be asymptotic. 
There are situations where the OPE cannot be applied or where it
cannot give definite answers. When $Q$ approaches $\Lambda$,
the series breaks down.
The OPE can also not answer questions related to the
asymptotics of the series due to the truncation of the series.   
In this talk I discuss a generalization of the local OPE that has been
proposed in Ref.\,\cite{us} and has been called "Delocalized
Operator Expansion" (DOE). The DOE is an attempt to
combine the advantages of the OPE with methods that might be useful to
resolve issues that cannot be tackled easily in the framework
of the OPE.

\section{BASIC FRAMEWORK}

\begin{figure}
\begin{center}
\leavevmode
\epsfxsize=4.5cm
\leavevmode
\epsffile[80 140 534 310]{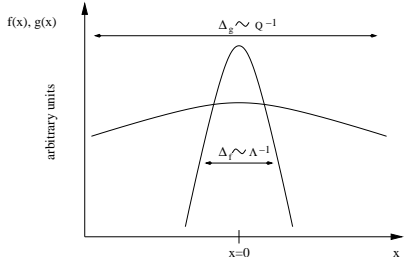}
%
\end{center}
\caption{\small Schematic picture of the short-distance function $f(x)$ and
the 2-point-correlator $g(x)$ illustrating the scale hierarchy
$\Delta_f/\Delta_g\sim\Lambda/Q\ll 1$.
\label{figfandg}
}   
\end{figure}

Given the exact knowledge of (nonlocal) gauge-invariant QCD vacuum
correlators and assuming for now that perturbative propagation can be
factorized unambiguously from nonperturbative fluctuations, one may
view the OPE as the multipole expansion of the perturbative
part. To see this, let us consider the simplified case of a
nonperturbative and nonlocal dimension-4 structure $g(x)$
corresponding to a {\sl slowly varying} 2-point correlator falling off
at distance $\Delta_g\sim \Lambda^{-1}$ in its Euclidean space-time
argument $x$. In addition, we consider a perturbative short-distance
function $f(x)$, that probes the vacuum at distance $x\sim\Delta_f\sim
Q^{-1}$. For $Q\gg\Lambda$, $f(x)$ is {\sl strongly peaked} compared
to $g(x)$ (see Fig.\,\ref{figfandg}). 
For illustration I consider the problem in one dimension. Then
the chain of local power corrections to the perturbative result is
obtained from the expression  
\begin{eqnarray}&& \textstyle
\label{mult}
\int_{-\infty}^\infty dx\,f(x)\,g(x) =: (f,g)
\,,
\end{eqnarray}
which is the bilinear form in the dual
space spanned by strongly peaked functions such as $f$ and
slowly varying functions such as $g$. 
The generalization to more than
one dimension, which also allows the treatment of non-perturbative
$n$-point correlators, is straightforward. 
In the OPE $f(x)$ is expanded in analogy to the
multipole expansion of a localized charge distribution in
electrostatics. This can be formulated by introducing
the dual space basis functions ($n=0,\ldots,\infty$)
\begin{eqnarray}&&\textstyle
e_n(x) \equiv \frac{(-1)^n}{n!}\,\delta^{(n)}(x)\,,\quad
\tilde e_n(x) \equiv  x^n
\,,
\label{deltabasis}
\end{eqnarray}
with the orthonormality relation $(e_n,\tilde e_m)=\delta_{nm}$.
The multipole expansion of $f(x)$ then reads
\begin{eqnarray}&&  \textstyle
f(x) = 
\sum_{n=0}^{\infty}
(f,\tilde e_n)e_n(x)
\end{eqnarray}
which leads to
\begin{eqnarray}
\label{fglocal}
\int\!\!\!dx f(x)g(x)= 
\sum_{n=0}^{\infty}
(f,\tilde e_n)(e_n,g)
=: \sum_{n=0}^{\infty} \! f_n g_n.
\end{eqnarray}
The $f_n$'s are the Wilson coefficients and the $g_n$ the matrix
elements of operators obtained from locally expanding the  
field content in the correlator $g(x)$ for small $x$.
In momentum space representation the Wilson coefficients and matrix
elements have the generic form
\begin{eqnarray} \textstyle
f_n = (i\frac{d}{dk})^n\,\tilde f(k)|_{k=0}\,,\,\,
g_n =\int\frac{dk}{2\pi}\,\frac{(ik)^n}{n!}\,\tilde g(k)\,,
\label{fngndeflocal}
\end{eqnarray}
where $\tilde f$ and $\tilde g$ are the Fourier transforms of $f$ and
$g$, respectively.

Briefly switching to 4 dimensions, a simple example for the
function $g(x)$ is the gauge invariant field strength
correlator~\cite{Dosch1} 
\begin{equation}
\label{gfc}
g_{\tiny \mu\nu\kappa\lambda}(x) \!\equiv\! 
\mbox{Tr}\langle g^2 G_{\mu\nu}(x)S(x,0)
G_{\kappa\lambda}(0)S(0,x)\rangle ,
\end{equation}
where
\begin{equation} \textstyle
S(x,y) 
\, = \,
{\cal P}\,\exp\{ig\int_0^x dz_\mu A_\mu(z)\}
\,.
\label{gaugestring}
\end{equation}
It generates the following chain of vacuum expectation values (VEV's)
involving local operators of increasing dimension
\begin{eqnarray}
\label{gluoncondensate}
\mbox{Tr}\langle g^2 G_{\mu\nu}G_{\kappa\lambda}\rangle,
\mbox{Tr}\langle g^2 G_{\mu\nu}D_\rho G_{\kappa\lambda}\rangle,
\nonumber\\
\mbox{Tr}\langle g^2 G_{\mu\nu}D_\tau D_\rho G_{\kappa\lambda}\rangle\,,
\ldots\,.\mbox{\hspace{8mm}}
\end{eqnarray}
In the limit, where $R\equiv\Delta g/\Delta f\to\infty$, 
the VEV with the lowest dimension, usually called the gluon
condensate, dominates, and the contributions of higher dimensional
VEV's are suppressed by higher powers of $\Lambda/Q$. VEV's with odd numbers of covariant
derivatives do not contribute due to parity and time-reversal
invariance. Perturbatively, one can define the 
gluon condensate in such a way that its anomalous dimension vanishes
to all orders in $\alpha_s$.

Improved convergence properties may be achieved using
a multipole expansion of $f(x)$ based on functions of width
$\Delta_f\approx Q^{-1}$ instead of the infinitely narrow
$\delta$-functions. {\it This is the basic idea in the construction of
the DOE}. At this point I would like to note that there exist a number of 
phenomenological studies where the non-local expression in
Eq.\,(\ref{mult}) has been analyzed directly without any expansion for
cases where the OPE did not have a good convergence behavior, see
e.g.\,Refs.\,\cite{nonlocal}. This
approach has the feature that it requires a model-dependent ansatz
for the correlation function $g(x)$ and that the computation of $f(x)$
can become cumbersome, particularly at higher loop level. 
The DOE has been constructed with the aim to provide an
alternative formalism to describe non-local effects. As we will see
later the DOE simplifies numerical predictions in a given model for
$g(x)$. However, I will also show that the DOE allows to extract
non-local non-perturbative information on the QCD vacuum in a
model-independent way.  

Let me continue with the construction of the DOE.
For Cartesian coordinates the dual space basis 
\begin{eqnarray}&&\textstyle
e_n^\Omega(x) \equiv
\frac{\Omega^{n+1}}{\sqrt{\pi}\,n!}H_n(\Omega x)e^{-\Omega^2 x^2}
\,,
\nonumber  \\&& \textstyle
\tilde e_n^\Omega(x) \equiv \frac{H_n(\Omega x)}{(2 \, \Omega)^n}
\,,
\label{hermitebasis}
\end{eqnarray}
with $(e_n^\Omega,\tilde e_n^\Omega)=\delta_{mn}$,
the $H_n$ being the Hermite polynomials,  is well suited.
Of course this choice is not unique, but it fixes a scheme, which can
be unambiguously related to other  
possible schemes that can be used.  
The $e_n^\Omega$ have a width of order $\Omega^{-1}$, and for
$\Omega\to\infty$ one finds 
$e_n^\Omega\to e_n$,
$\tilde e_n^\Omega\to\tilde
e_n$. The parameter $\Omega$ is called {\em resolution scale}.
In this basis Eq.\,(\ref{mult}) can be written as  
\begin{eqnarray} 
\label{fgdelocalized}  \textstyle
\sum_{n=0}^{\infty}
(f,\tilde e_n^\Omega)(e_n^\Omega,g)
=: \sum_{n=0}^{\infty} \! f_n(\Omega) g_n(\Omega),
\end{eqnarray}
where the $\Omega$-dependent short-distance coefficients and matrix
elements have the form 
\begin{eqnarray}&&\textstyle
f_n(\Omega)=
\frac{1}{(2\Omega)^n}H_n(\Omega(i\frac{d}{dk}))\tilde f(k)|_{k=0}\,,
\nonumber\\&&\textstyle
g_n(\Omega)=
\int\frac{dk}{2\pi}\frac{(i k)^n}{n!}
e^{-\frac{k^2}{4\Omega^2}}\tilde g(k)\,,
\label{fngndefdelocal}
\end{eqnarray}
in momentum space representation. 
The series in Eq.\,(\ref{fgdelocalized}) 
has better convergence properties, if $\Omega$ is of order $Q$ rather
than being equal to $\infty$, because the first term in the
delocalized multipole expansion generally provides a better
approximation of the actual form of $f(x)$ than the local expansion.
The relation between basis functions for resolution parameters
$\Omega$ and $\Omega^\prime$ reads
\begin{eqnarray}&&\textstyle
\tilde e_n^\Omega(x) =
\sum_{m=0}^{\infty} a_{nm}(\Omega,\Omega^\prime)\,
\tilde e_m^{\Omega^\prime}(x)
\nonumber\\&&\textstyle
e_n^\Omega(x) =
\sum_{m=0}^{\infty} a_{mn}(\Omega^\prime,\Omega)\,
e_m^{\Omega^\prime}(x)
\,,
\label{changeofbasis}
\end{eqnarray}
where
\begin{equation}
a_{nm}(\Omega,\Omega^\prime)=
\frac{n!}{m!(\frac{n-m}{2})!}\,
\bigg({\displaystyle \frac{\Omega^2-\Omega^{\prime\,2}}
      {4\,\Omega^2\,\Omega^{\prime\,2}}}\bigg)^{\big(\mbox{$\frac{n-m}{2}$}\big)}
\label{anmdef}
\end{equation}
if $n-m \ge 0$ and even, and $a_{nm}=0$ otherwise.
The transformations  $a_{nm}(\Omega,\Omega^\prime)$ form a
group. One can relate the short-distance coefficients and matrix
elements for different resolution parameters to each other. Here, I
only want to discuss these relations for finite $\Omega$ and
$\Omega^\prime=\infty$: 
\begin{eqnarray}
\label{fgrelationOmegainfty}
&&\textstyle
f_0(\Omega)=f_0\,, 
\\&&\textstyle
f_2(\Omega)=f_2-\frac{1}{2\Omega^2} f_0\,,\ldots ,
\nonumber\\&&\textstyle
f_n(\Omega)=\sum_{i=0}^{[n/2]}\frac{n!}{(n-2i)! i!}
	    (-\frac{1}{4\Omega^2})^i\,f_{n-2i}\,,
\nonumber\\&&\textstyle
g_n(\Omega)=
\sum_{i=n}^\infty\frac{(n+2i)!}{n! i!}
\Big(\frac{1}{4\Omega^2}\Big)^i g_{n+2i}
\,.
\label{gndoeope}
\end{eqnarray}
The coefficient $f_n(\Omega)$ can be
expressed in terms of a {\it finite} 
linear combination of the local Wilson coefficients $f_i(\infty)$ for
$i\le n$. The short-distance coefficient of the leading
power correction is $\Omega$-independent. The
$\Omega$-dependent matrix elements $g_n(\Omega)$ are related to an
{\it infinite} sum of local matrix elements with additional covariant
derivatives. These properties exist for all dual space bases
constructed from orthogonal polynomials. Note that
relations\,(\ref{fgrelationOmegainfty}) can also be employed if
the separation between long- and short-distance
contributions is factorization scale dependent, and one can therefore use
these relations as the formal definition of the terms in the DOE.

The DOE of quantities such as in Eq.\,(\ref{Wilson}) has the same
parametric counting in powers of $\Lambda/Q$ as the OPE, if
$\Omega$ is not chosen parametrically smaller than $Q$. Consider that
$g_n\sim\Lambda^n$ and $f_n\sim Q^{-n}$ in the OPE, then we have
\begin{eqnarray}&&\textstyle
g_n(\Omega)\sim \Lambda^n \sum_i (\frac{\Lambda}{\Omega})^i \sim
\Lambda^n
\,,
\nonumber\\&&\textstyle
f_n(\Omega)\sim Q^{-n} \sum_i (\frac{Q}{\Omega})^i \sim Q^{-n}
\,,
\end{eqnarray}
as long as $\Omega\gsim Q$. However, one expects that the actual size of the
term $f_n(\Omega)g_n(\Omega)$ in the DOE for $\Omega\sim Q$ has an
additional numerical suppression by powers of a small number.
 
\section{HEAVY QUARKONIUM GROUND STATE ENERGY}
\label{sectionenergy}

\begin{figure}
\begin{center}
\leavevmode
\epsfxsize=2cm
\epsffile[260 500 420 700]{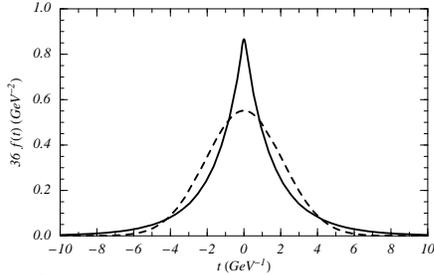}
%
%
\vspace{-0.2cm}
 \caption{\label{figffunction} \small
The perturbative short-distance function $f(t)$ for $m=5$~GeV and
$\alpha_s=0.39$ (solid line) and the leading term in the delocalized
multipole expansion of $f(t)$, 
$[\int dt^\prime f(t^\prime)\tilde e_0^\Omega(t^\prime)]$
for $\Omega=k^2/m$ 
(dashed line). 
}
 \end{center}
\end{figure}

In heavy quarkonium
systems the relevant physical scales, mass $m$, momentum $p$,
energy $E$ and $\Lambda$ have the hierarchy
\begin{eqnarray}&&
m \gg p\sim m v \gg E\sim m v^2 \gg \Lambda
\,,
\end{eqnarray}
where $v\ll 1$ is the quark velocity.  
Thus the spatial size $\sim (m v)^{-1}$ is
much smaller than the typical dynamical time scale $\sim (m
v^2)^{-1}$. 
In this section I demonstrate the DOE in a toy-model computation for
the nonperturbative corrections to the $1^{\,3}S_1$ ground 
state for the expansion in $\Lambda/E$. My intention is not to carry
out a phenomenological study, but to show how well the series in the
DOE behaves in comparison to the series in OPE and how well the DOE
approximates the exact model result.
The ratios of 
$m$, $p$ and $E$ are treated at leading order in the local
expansion. This means that the perturbative dynamics is
described by the nonrelativistic two-body Schr\"odinger
equation and that the interaction with the nonperturbative vacuum 
is accounted for by two insertions of the local 
$\bmx\bmE$ dipole operator, $\bmE$ being the chromoelectric
field.\,\cite{Voloshin1} The chain of VEV's of the two gluon operator
with increasing numbers of covariant derivatives 
times powers of quark-antiquark octet propagators~\cite{Voloshin1},
is treated in the DOE. In this model the interaction with the
vacuum fluctuations only depends on the temporal distance (in
Euclidean space) of two insertions of $\bmx\bmE$ dipole operators and
the nonperturbative correction to the ground state energy reads
($k=\frac{2}{3}m\alpha_s$)
\begin{eqnarray}&&\textstyle
E^{np} = \int_{-\infty}^\infty\! dt \, f(t) \, g(t)
\,,
\label{Enonpertdef}
\nonumber\\&&\textstyle
f(t)=
\int \frac{dq_0}{72\pi}e^{-q_0 t}
\int\! d^3\bmx \int\! d^3\bmy
\phi(x)(\bmx\bmy)
\nonumber\\&&\textstyle \qquad\qquad\times
G_o(\bmx,\bmy,-\frac{k^2}{m}-q_0)\,\phi(y)
\,,
\label{Enonpertfdef}
\end{eqnarray}
where $f(t)$ is the short-distance function 
that depends on ground state Coulomb wave function $\phi$ and the
octet Green-function $G_o$~\cite{Voloshin1} and $g(t)$ is the
gluon field strength correlator for $x=(t,0,0,0)$.
In Fig.\,\ref{figffunction} the function $f(t)$ (solid line) is
displayed for $m=5$\,GeV and $\alpha_s=0.39$.
The characteristic width of $f$ is of order the energy  
$k^2/m\sim (m \alpha_s^2)^{-1}\sim(m v^2)^{-1}$. For the nonperturbative gluonic
field strength correlator I use a lattice-inspired~\cite{Delia} model  
\begin{eqnarray}&&\textstyle
g(t)=12\,A_0\exp(-\sqrt{t^2+\lambda_A^{2}}/\lambda_A + 1)
\nonumber\\&&\textstyle
A_0=0.04~\mbox{GeV}^{4}\,,\qquad
\lambda_A^{-1}=0.7~\mbox{GeV}\,,
\label{gmodel}
\end{eqnarray}
with a large-time behavior $\sim e^{-t/\Lambda_A}$. In
this model the value of the gluon condensate is
$\langle\frac{\alpha_s}{\pi}\,G_{\mu\nu}^a G_{\mu\nu}^a\rangle
=\frac{6\,A_0}{\pi^2}=0.024\,\,\mbox{GeV}^4$.
\begin{table}[t!]  
\begin{center}
\begin{small}
\begin{tabular}{|c|c|c||r|r|r|} \hline
\multicolumn{4}{|c|}{} & \multicolumn{1}{|c|}{$\Omega=\infty$} 
                  & \multicolumn{1}{|c|}{$\Omega=\frac{k^2}{m}$}
 \\ \hline
  & $k^2/m$ & $E^{np}$ & & $f_ng_n$  
                              & $f_ng_n$
 \\ 
 \raisebox{1.5ex}[-1.5ex]{$\alpha_s$} & (GeV) & (MeV)
   & \raisebox{1.5ex}[-1.5ex]{$n$} & (MeV) & (MeV) 
 \\ \hline\hline
   $0.39$ & $0.338$ & $24.8$ & 
           0 & $38.6$      & $24.2$  
\\ \cline{4-6}
& & &    2 & $-65.7$     & $-3.9$ 
\\ \cline{4-6}
& & &    4 & $832.7$     & $12.1$ 
\\ \cline{4-6}
& & &    6 & $-35048.0$  & $-43.1$ 
\\ \hline\hline
 $0.15$ & $1.750$ & $0.245$ & 
           0 & $0.258$   &  $0.249$  
\\ \cline{4-6}
& & &    2 & $-0.016$  &  $-0.005$  
\\ \cline{4-6}
& & &    4 & $0.008$   &  $0.003$   
\\ \cline{4-6}
& & &    6 & $-0.012$  &  $-0.003$  
\\ \hline
\end{tabular}
\caption{\label{tabquarkonium} \small
Nonperturbative corrections to the heavy quarkonium ground state level
as described in the text.
}
\end{small}
\end{center}
\end{table}
In Tab.\,\ref{tabquarkonium} the exact result and
the first four terms of the resolution-dependent expansion of $E^{np}$
are shown for the quark masses $m=5$ (upper part) and $175$~GeV (lower
part) and for $\Omega=\infty$ and $\Omega=k^2/m$. The strong coupling
has been fixed by the relation $\alpha_s=\alpha_s(k)$. The numerical
values of $f_n\,g_n$ have been determined from 
Eqs.\,(\ref{fngndeflocal},\ref{fngndefdelocal}).
The series are all asymptotic. The local expansion
is badly behaved for $m=5$~GeV because 
$k^2/m\lsim\lambda_A^{-1}$, and basically meaningless. 
For $m=175$~GeV, where $k^2/m>\lambda_A^{-1}$, the local
expansion is good. For $\Omega=k^2/m$, however, the series
is much better behaved for all quark masses. The size of the 
order $n$ term is about a factor $2^{-n}$ smaller than
the order $n$ term in the local expansion. 
One also observes that even in the case $k^2/m < \lambda_A^{-1}$
the leading term in the delocalized expansion for
$\Omega=k^2/m$ agrees with the exact result within a few percent. This
feature is a general property of the delocalized expansion, and should
apply to any quantity for which the local expansion in the ratio of two
scales breaks down because the ratio is not sufficiently small.

\section{RUNNING GLUON CONDENSATE FROM CHARMONIUM SUM RULES}

\begin{figure}[t!] 
\begin{center}
\leavevmode
\epsfxsize=2cm
\epsffile[260 500 420 700]{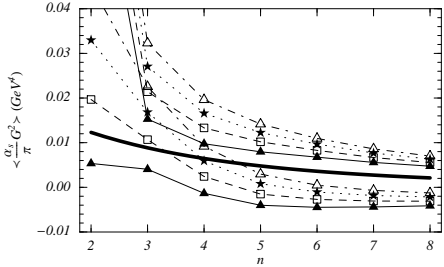}
%
%
\vspace{-0.2cm}
 \caption{\label{figcharmonium}
\small 
The running gluon condensate as a function of $n$ when extracted from
the ratio of charmonium moments $r_n$.
}
 \end{center}
\end{figure}

The $\Omega$-dependent matrix elements are either determined 
from experimental data or from lattice
measurements. In the following the $\Omega$-dependent (``running'')
gluon condensate is extracted  from charmonium sum rules
which are based on moments~\cite{Novikov1}
\begin{eqnarray}&&\textstyle
{\cal M}_n =
\frac{1}{n!} (-\frac{d}{dQ^2})^n
\Pi^c(Q^2)|_{Q^2=0} 
\label{momdef1}
\end{eqnarray}
of the correlator of two charm quark vector currents
$j_\mu\equiv\bar{c}\gamma_\mu c$. The moments can be determined
theoretically in an expansion of the form~\cite{Nikolaev1}
\begin{equation}\textstyle
{\cal M}_n =
{\cal M}_n^0
\{ 1+\mbox{[p. corr.]} +
\delta^{(4)}_n\,\langle g^2 G^2\rangle + \ldots\}
\,,
\label{Mntheodef}
\end{equation}
while the experimental moments are obtained from a dispersion integral
over the $c\bar c$ cross section in $e^+e^-$ annihilation.
The Wilson coefficient of the gluon condensate,
$\delta^{(4)}_n$, is $\Omega$-independent.
We consider the ratio~\cite{Novikov1}
$
r_n 
\equiv
\frac{{\cal M}_n}{{\cal M}_{n-1}}
$
and extract the gluon condensate as a function of $n$.
Since the relevant short-distance scale for the moment ${\cal M}_n$
is of order $m_c/n$, a proper choice for the resolution scale is
$\Omega=2m_c/n$. Thus the dependence of the gluon condensate on $n$
can be interpreted as the dependence on $\Omega$. 
The results for the gluon condensate as a function of
$n$ is shown in Fig.\,\ref{figcharmonium} for $n\le 8$ and 
$\overline{m}_c(\overline{m}_c)=1.23$ (white triangles), $1.24$ 
(black stars), $1.25$ (white squares) and $1.26$~GeV (black
triangles). The area between the upper and lower symbols indicates
the experimental and theoretical uncertainties.
The running gluon condensate appears to be a decreasing
function of $n$. The solid thick line is 
$\langle\frac{\alpha_s}{\pi}\,G^2\rangle(\Omega)$
obtained from the form of the gluon field strength correlator
suggested from lattice computations~\cite{Delia}
for $\Omega=(2.5\,\,\mbox{GeV})/n$ and the vacuum correlation length  
$\lambda_A^{-1}=0.7$~GeV. 
The qualitative agreement is encouraging 
but, the uncertainties of our extraction are
still quite large. (See Ref.\,\cite{us} for an analysis
based on hadronic $\tau$ decay data.)

\section{CONCLUSIONS}

In this talk I discussed a generalization of the Wilson OPE
based on nonlocal projections of gauge invariant correlation
functions in a delocalized version of the multipole expansion
for the perturbatively calculable coefficient functions.~\cite{us} 
This "Delocalized Operator Expansion" (DOE) depends on an additional
parameter $\Omega$, called "resolution scale" which adjusts the width
of the projection functions used for the expansion. The DOE has 
the same power counting as the OPE, but in general better convergence
properties than the OPE. The DOE 
short-distance coefficients can be determined from the 
OPE Wilson coefficients at the same order in the expansion,
whereas the DOE matrix elements correspond to an infinite sum of OPE
matrix elements with additional covariant derivatives. I believe that
the DOE can serve as a useful tool for situations where the OPE cannot be
applied.

\end{document}